\documentclass[aps,pra,twocolumn,groupedaddress,showpacs]{revtex4}

\usepackage{amsmath}

\begin{document}
%\title{Generalized Schwinger Representation with Fermions}
\title{Non-standard Schwinger fermionic representation of unitary group}
\author{Fu-Lin Zhang}
%\email[Email:]{flzhang@mail.nankai.edu.cn}
\affiliation{Theoretical Physics Division, Chern Institute of
Mathematics, Nankai University, Tianjin 300071, P.R.China}
\author{Jing-Ling Chen}
\email[Email:]{chenjl@nankai.edu.cn} \affiliation{Theoretical
Physics Division, Chern Institute of Mathematics, Nankai University,
Tianjin 300071, P.R.China}

\date{\today}

\begin{abstract}
The non-standard Schwinger fermionic representation of the unitary
group is studied by using $n$-fermion operators. One finds that the
Schwinger fermionic representation of the $U(n)$ group is not unique
when $n\ge 3$. In general, based on $n$-fermion operators, the
non-standard Schwinger fermionic representation of the $U(n)$ group
can be established in a uniform approach, where all the generators
commute with the total number operators. The Schwinger fermionic
representation of  $U(C^{m}_{n})$ group is also discussed.
\end{abstract}

%\keywords{}

\maketitle

\section{Introduction \label{intro}}

There are many kinds of important mappings between spin systems and
multi-boson or multi-fermion systems. They not only essentially
simplify the analysis of the problems under consideration, but also
help us to understand various aspects of them. The most famous among
these mappings are the Schwinger \cite{sch,sch1}, Holstein-Primakoff
\cite{HP} bosonic representations and the Jordan-Wigner
transformations \cite{JW}. The first two map an arbitrary spin to a
bosonic system, while the third estblishes a connection between
one-dimensional spin-$1/2$ lattice and spinless formions on the same
lattice. These representations are very successful in describing
magnetism in various quantum systems
\cite{Timm1,Timm2,LM1,LM2,AA1,AA2}. They also play a significant
role in many other contexts. For example, the Schwinger
representation has been exploited in quantum optics \cite{Arvind}
and in the study of certain classes of partially coherent optical
beams \cite{Simon}. The Jordan-Wigner transformation has also been
used to simulate the interacting fermions with quantum computers
\cite{Simulating}.

A lot of work has been done to generalize these mappings in several
ways. Chaturvedi and Mukunda have extended the Schwinger
representation to the $SU(3)$ case on the primes of that each
unitary irreducible representation appears exactly once \cite{su3}.
Kim has discussed the Schwinger representation based on the mixed
sets of creation and annihilation operators of bosonic and fermionic
type \cite{mixed}. The Jordan-Wigner transformation has been
generalized to arbitrary spin case and also the fermions were
replaced by anyons \cite{JWG,JWF}. In addition, there are some new
kinds of constructions of the $SU(2)$ algebra being reported
\cite{DX,SDX}.

% The essential different between the
%Schwinger boson representation and the Jordan-Wigner transformation
%is that the phase space of the bosonic system is
%infinite-dimensional but the one of the fermionic is
%finite-dimensional. So the Jordan-Wigner transformation is
%invertible without requiring constraint but the Schwinger
%representation not. In this paper, our aim is to generalize the
%Schwinger representation to the other Lie groups using fermionic
%operators.

The purpose of this Brief Report is to study the Schwinger
representation of unitary group by using $n$-fermion operators. The
work is organized as follows: In Sec. II, to make the report be
self-contained, we make a brief review for the standard Schwinger
representation (SSR) of unitary group with bosons and fermions. In
Sec. III, we show that the realization of the $U(3)$ group with
3-fermion is not unique, by providing a non-standard Schwinger
fermionic representation (NSSFR). In Sec. IV, we develop a uniform
approach to establish the NSSFR of the $U(n)$ group based on
$n$-fermion operators ($n\ge 3$), where all the generators commute
with the total number operators. In Sec. V, we discuss the Schwinger
fermionic representation of the $U(C^{m}_{n})$ group based on $n$
fermions. Conclusion is made in the last section.

%In the present paper, we hold on the common point of the classical
%Schwinger boson and fermion representations that the generators are
%commutable with the total number operators, and discuss the more
%universal constructions of the unitary groups using fermionic
%operators. In the section \ref{un}, we will introduce a new type
%fermion representation of the $SU(3)$ and give the $n$-fermion
%representations of the $SU(n)$ sequentially. A farther consequence
%in the section \ref{ucnm} will show the scheme to construct the
%$SU(C^{m}_{n})$ Lie algebra using $n$ fermionic operators. The
%section \ref{conclu} is the concluding remarks.

\section{Brief review of SSR}

% Let us give some briefly review.
In the standard Schwinger bonsonic representation of $SU(2)$ group,
the three generators are mapped onto the bilinear form of the
bosonic operators as
%\begin{eqnarray}\label{su2b}
%J^{b}_{1}&=&(b^{\dag}_{1}b_{2}+b^{\dag}_{2}b_{1})/2, \nonumber\\
%J^{b}_{2}&=&(b^{\dag}_{1}b_{2}-b^{\dag}_{2}b_{1})/2i, \\
%J^{b}_{3}&=&(b^{\dag}_{1}b_{1}-b^{\dag}_{2}b_{2})/2 \nonumber,
%\end{eqnarray}
\begin{eqnarray}\label{su2b}
J^{b}_{1}&=&\frac{1}{2}(b^{\dag}_{1}b_{2}+b^{\dag}_{2}b_{1})\nonumber\\
J^{b}_{2}&=&\frac{1}{2i}(b^{\dag}_{1}b_{2}-b^{\dag}_{2}b_{1})\\
J^{b}_{3}&=&\frac{1}{2}(b^{\dag}_{1}b_{1}-b^{\dag}_{2}b_{2})\nonumber,
\end{eqnarray}
where $b_i$ and $b^\dag_i$ ($i$=1,2) are the annihilation and
creation operators of the $i$-th boson, respectively.

The commutation relations of $n$ independent bosonic operators are
\begin{eqnarray}\label{boson}
[b_i,b^\dag_j]=\delta _{ij},\ [b_i,b_j]=[b^\dag_i,b^\dag_j]=0.
\end{eqnarray}
Let $Q^{b}_{ij}=b^\dag_i b_j$, it is easy to have
\begin{eqnarray}\label{boson2}
[Q^{b}_{ij},Q^{b}_{kl}]=\delta_{jk} Q^{b}_{il}-\delta_{li}
Q^{b}_{kj},
\end{eqnarray}
which remarkably shares the same commutation relation of matrices
$e_{ij}$, having $1$ in the $(i, j)$ position (i.e., the $i$-th
arrow and the $j$-th column of the matric) and $0$ elsewhere. If
$G_{i}$ is a matrix generator of a Lie group and $G^{\alpha
\beta}_{i}$ denotes its elements in the $(\alpha, \beta)$ position,
then the operators $G^{b}_{i}$, which are defined by the linear
combinations of $Q^{b}_{ij}$ as
\begin{eqnarray}\label{schrep}
G^{b}_{i}=\sum_{\alpha \beta} G^{\alpha \beta}_{i}Q^{b}_{\alpha
\beta}=\sum_{\alpha \beta} b^\dag_{\alpha} G^{\alpha \beta}_{i}
b_{\beta},
\end{eqnarray}
obey the same commutation relations as those of the matrices
$G_{i}$. It means that the operators $G^{b}_{i}$ form a
representation of the Lie group represented by $G_{i}$. When $n=2$
and $G_{i}=\sigma_{i}/2$ (where $\sigma_{i}$ are the Pauli matrices,
$i=1, 2, 3$), Eq. (\ref{schrep}) gives the standard Schwinger
bosonic representation of the $U(2)$ group as in Eq. (\ref{su2b}).

The annihilation and creation operators in Eq. (\ref{schrep}) are
not necessarily restricted to bosons. Notice that the
anticommutation relations of $n$ independent fermions are
\begin{eqnarray}\label{fermion}
\{a_i,a^\dag_j\}=\delta _{ij},\ \{a_i,a_j\}=\{a^\dag_i,a^\dag_j\}=0,
\end{eqnarray}
where $a_i$ and $a^\dag_i$ are annihilation and creation operators
of the $i$-th fermion. The key point is that the bilinear form
operators $Q^{f}_{ij}=a^{\dag}_{i}a_{j}$ also obey the commutation
relations of $e_{ij}$. Thus replacing $Q^{b}_{\alpha \beta}$ by
$Q^{f}_{\alpha \beta}$ in Eq. (\ref{schrep}), one gets the standard
Schwinger fermionic representation of the Lie group
\begin{eqnarray}\label{schrepf}
G^{f}_{i}=\sum_{\alpha \beta} G^{\alpha \beta}_{i}Q^{f}_{\alpha
\beta}=\sum_{\alpha \beta} a^\dag_{\alpha} G^{\alpha \beta}_{i}
a_{\beta}.
\end{eqnarray}

There are $n^2$ independent operators $Q^{b}_{ij}$ or $Q^{f}_{ij}$
for $n$ bosons or fermions. They can construct the $U(n)$ Lie
algebra, and all the generators commute with the total number
operator
$N^{b}=\sum^{n}_{i=1}N^{b}_{i}=\sum^{n}_{i=1}b^{\dag}_{i}b_{i}$, or
$N^{f}=\sum^{n}_{i=1}N^{f}_{i}=\sum^{n}_{i=1}a^{\dag}_{i}a_{i}$.
%\begin{eqnarray}\label{Noper}
%N^{b}=\sum^{n}_{i=1}N^{b}_{i}=\sum^{n}_{i=1}b^{\dag}_{i}b_{i}\\
%N^{f}=\sum^{n}_{i=1}N^{f}_{i}=\sum^{n}_{i=1}a^{\dag}_{i}a_{i}.
%\end{eqnarray}
Since the Hamiltonians of the $n$-dimensional bosonic and fermionic
oscillators are $H^{b}=N^{b}+\frac{n}{2}$,
$H^{f}=N^{f}+\frac{n}{2}$,
%\begin{eqnarray}\label{H}
%H^{b}=N^{b}+\frac{n}{2},\ H^{f}=N^{f}+\frac{n}{2}
%H^{b}=N^{b}+\frac{n}{2}=\sum^{n}_{i=1} N^{b}_{i}+\frac{n}{2}=\sum^{n}_{i=1}b^\dag_i b_i+\frac{n}{2}\\
%H^{f}=N^{f}+\frac{n}{2}=\sum^{n}_{i=1}
%N^{f}_{i}-\frac{n}{2}=\sum^{n}_{i=1}a^\dag_i a_i-\frac{n}{2}
%\end{eqnarray}
It indicates that both $n$-dimensional bosonic and fermionic
oscillators have the $U(n)$ symmetry.

\section{NSSFR of $U(3)$ group with 3-Fermion \label{un}}
In this section we would like to study the NSSFR of $U(3)$ group
with three fermions.
%In a representation, the orders of the
%annihilation and creation operators that commute with the total
%number operator are required equal.
For three fermions, there is only one independent $6$-order operator
$a^{\dag}_{1}a_{1}a^{\dag}_{2}a_{2}a^{\dag}_{3}a_{3}$ and no
higher-order operator. The 2-order operators have been used to
realized the standard Schwinger fermion representation. Here we
attend to apply the $4$-order operators to establish the NSSFR.

In the fundamental representation \cite{GM} of the $U(3)$, the
generators are $G_{i}=\lambda_{i}/2$, $i=1,2,...,8$, where
$\lambda_{i}$ are the Gell-Mann matrices. They obey the $U(3)$
algebraic commutation relations
\begin{eqnarray}\label{su3com}
[\lambda_{i},\lambda_{j}]=2i \sum^{8}_{k=1}f_{ijk} \lambda_{k},
\end{eqnarray}
where $f_{ijk}$ is the structure constant. Notice that the $j=1$
representations of the $U(2)$ algebra are
\begin{eqnarray}
J^{(1)}_{+}&=&J^{(1)}_{1}+iJ^{(1)}_{2}= \begin{bmatrix}
 0&\sqrt{2}&0\\
 0&0&\sqrt{2}\\
 0&0&0
\end{bmatrix},\nonumber \\
\ J^{(1)}_{-}&=&J^{(1)}_{1}-iJ^{(1)}_{2}= \begin{bmatrix}
 0&0&0\\
 \sqrt{2}&0&0\\
 0&\sqrt{2}&0
\end{bmatrix},\\
\ J^{(1)}_{3}&=& \begin{bmatrix}
 1&0&0\\
 0&0&0\\
 0&0&-1
\end{bmatrix}.\nonumber
\end{eqnarray}
The Gell-Mann matrices can be constructed by using the quadratic
forms of the above spin-$1$ matrices as
\begin{eqnarray}\label{spin1togellmann}
\lambda_{1}&=&\frac{\sqrt{2}}{2}(J^{(1)}_{3}J^{(1)}_{+}+J^{(1)}_{-}J^{(1)}_{3}),\nonumber \\
\lambda_{2}&=&\frac{-i\sqrt{2}}{2}(J^{(1)}_{3}J^{(1)}_{+}-J^{(1)}_{-}J^{(1)}_{3}),\nonumber \\
\lambda_{3}&=&\frac{1}{2}[J^{(1)}_{3}J^{(1)}_{+},J^{(1)}_{-}J^{(1)}_{3}],\nonumber \\
\lambda_{4}&=&\frac{1}{2}({J^{(1)}_{+}}^2+{J^{(1)}_{-}}^2),\nonumber \\
\lambda_{5}&=&\frac{-i}{2}({J^{(1)}_{+}}^2-{J^{(1)}_{-}}^2),\\
\lambda_{6}&=&\frac{-\sqrt{2}}{2}(J^{(1)}_{3}J^{(1)}_{-}+J^{(1)}_{+}J^{(1)}_{3}),\nonumber \\
\lambda_{7}&=&\frac{i\sqrt{2}}{2}(J^{(1)}_{3}J^{(1)}_{-}-J^{(1)}_{+}J^{(1)}_{3}),\nonumber \\
\lambda_{8}&=&\frac{1}{4\sqrt{3}}[{J^{(1)}_{+}}^2,{J^{(1)}_{-}}^2]+\frac{1}{2\sqrt{3}}[J^{(1)}_{3}J^{(1)}_{-},J^{(1)}_{+}J^{(1)}_{3}].\nonumber
\end{eqnarray}

Conseuently, if the spin-$1$ matrices in Eq. (\ref{spin1togellmann})
are replaced by their standard Schwinger fermionic representations
\begin{eqnarray}
J^{f}_{+}&=&\sqrt{2}(a^{\dag}_{1}a_{2}+a^{\dag}_{2}a_{3}),\nonumber\\
J^{f}_{-}&=&\sqrt{2}(a^{\dag}_{2}a_{1}+a^{\dag}_{3}a_{2}),\\
J^{f}_{3}&=&a^{\dag}_{1}a_{1}-a^{\dag}_{3}a_{3}=N_{1}-N_{3},\nonumber
\end{eqnarray}
a higher-order representation of the $U(3)$ group can then be
obtained as
\begin{eqnarray}\label{hordersu3}
\lambda^{h}_{1}&=&(a^{\dag}_{1}a_{2}+a^{\dag}_{2}a_{1})(1-N_{3})+(a^{\dag}_{2}a_{3}+a^{\dag}_{3}a_{2})N_{1},\nonumber \\
\lambda^{h}_{2}&=&(-i a^{\dag}_{1}a_{2}+i a^{\dag}_{2}a_{1})(1-N_{3})+(-i a^{\dag}_{2}a_{3}+i a^{\dag}_{3}a_{2})N_{1},\nonumber \\
\lambda^{h}_{3}&=&N_{1}-N_{2}-2N_{1}N_{3}+N_{1}N_{2}+N_{2}N_{3},\nonumber \\
\lambda^{h}_{4}&=&(a^{\dag}_{1}a_{3}+a^{\dag}_{3}a_{1})(1-2N_{2}),\nonumber \\
\lambda^{h}_{5}&=&(-i a^{\dag}_{1}a_{3}+i a^{\dag}_{3}a_{1})(1-2N_{2}),\\
\lambda^{h}_{6}&=&(a^{\dag}_{2}a_{3}+a^{\dag}_{3}a_{2})(1-N_{1})+(a^{\dag}_{1}a_{2}+a^{\dag}_{2}a_{1})N_{3},\nonumber \\
\lambda^{h}_{7}&=&(-i a^{\dag}_{2}a_{3}+i a^{\dag}_{3}a_{2})(1-N_{1})+(-i a^{\dag}_{1}a_{2}+ia^{\dag}_{2}a_{1})N_{3},\nonumber \\
\lambda^{h}_{8}&=&(N_{1}+N_{2}-2N_{3}+2N_{1}N_{3}-N_{2}N_{3}-N_{1}N_{2})/\sqrt{3}.\nonumber
\end{eqnarray}
One may verify that the commutation relations of $\lambda^{h}_{i}$
are the same as in Eq. (\ref{su3com}) due to Eq. (\ref{fermion}).
All of the operators $\lambda^{h}_{i}$ commute with the total number
operator. It is worthy to mention that the non-standard Schwinger
representation in Eq. (\ref{hordersu3}) does not valid for bosons.

\section{NSSFR with $n$-fermion }

The above method of getting the NSSFR of $U(3)$ group is not
conveniently applicable for the arbitrary $U(n)$ group. Thus we
would like to rewrite Eq. (\ref{hordersu3}) into a uniform form, so
that it can be directly generalized to an arbitrary $U(n)$ group.

Let us look at the corresponding matrix-representation of Eq.
(\ref{hordersu3}) in the occupation number space, whose standard
basis reads
\begin{eqnarray}\label{basis}
\{1,\ a^{\dag}_{1},\ a^{\dag}_{2},\ a^{\dag}_{3},\
a^{\dag}_{1}a^{\dag}_{2},\ a^{\dag}_{1}a^{\dag}_{3},\
a^{\dag}_{2}a^{\dag}_{3},\ a^{\dag}_{1}a^{\dag}_{2}a^{\dag}_{3}\} \;
|vac\rangle,
%|\alpha\rangle=(a^{\dag}_{1})^{\alpha_{1}}(a^{\dag}_{2})^{\alpha_{2}}(a^{\dag}_{3})^{\alpha_{3}}|vac\rangle .
\end{eqnarray}
%where $\alpha_{1,2,3}=0$ or $1$.
where $|vac\rangle$ is the vacuum state. Based on Eq. (\ref{basis}),
the operators in Eq. (\ref{hordersu3}) correspond the following
$8\times 8$ partitioned matrices:
\begin{eqnarray}
\lambda^{h}_{i}=\begin{bmatrix}
 0&0&0&0\\
 0&\lambda_{i}&0&0\\
 0&0&\lambda_{i}&0\\
 0&0&0&0
\end{bmatrix},
\end{eqnarray}
where $\lambda_{i}$ denotes the $i$-th Gell-Mann matrix of $U(3)$.
Evidently, matrices $\lambda^{h}_{i}$ satisfy the commutation
relations of $U(3)$ algebra because matrices $\lambda_{i}$ do so.
Similarly, the standard Schwinger fermionic representation
$\lambda^{f}_{i}=\sum^{3,3}_{\alpha=1,\beta=1} a^\dag_{\alpha}
{\lambda_{i}}^{\alpha \beta} a_{\beta}$ correspond to the following
matrices
\begin{eqnarray}
\lambda^{f}_{i}=\begin{bmatrix}
 0&0&0&0\\
 0&\lambda_{i}&0&0\\
 0&0&\lambda^{\prime}_{i}&0\\
 0&0&0&0
\end{bmatrix},
\end{eqnarray}
where $\lambda^{\prime}_{i}=U(-\lambda^{*}_{i})U^{\dag}$,
$\lambda^{*}_{i}$ is the conjugate matrix of $\lambda_{i}$, and $U$
denotes the unitary matrix
\begin{eqnarray}
 U=\begin{bmatrix}
 0&0&1\\
 0&{-1}&0\\
 1&0&0
\end{bmatrix} .
\end{eqnarray}

In the $U(3)$ case, the occupation number space is divided into four
invariant subspaces specified by the total particle number, which
runs from 0 to $3$. Generally, the occupation number space of $U(n)$
group is divided into $n+1$ invariant subspaces, where the total
particle number $N$ runs from 0 to $n$, and the subspace with $N=m$
is conjugated to the subspace with $N=n-m$.

One may define a selective function
\begin{eqnarray}
f_{n}^{(m)}(x)=\prod^{m-1}_{i=1}\frac{x-i}{m-i}
\prod^{n-1}_{i=m+1}\frac{x-i}{m-i} .
\end{eqnarray}
$f_{n}^{(m)}(m)=1$, and $f_{n}^{(m)}(x)=0$ when $x$ equals to any
other integer between $1$ and $n-1$. Then the NSSFR of $U(3)$ in Eq.
(\ref{hordersu3}) can be recast to a very simple form
\begin{eqnarray}\label{fermu3}
\lambda^{h}_{i}&=&\sum^{n=3}_{\alpha,\beta=1} [ a^\dag_{\alpha}
{\lambda_{i}}^{\alpha \beta} a_{\beta} f_{3}^{(1)}(N) \nonumber\\&&+
 a^\dag_{\alpha}
{\lambda^{\prime}_{i}}^{\alpha \beta} a_{\beta} f_{3}^{(2)}(N)],
\end{eqnarray}
where $N=\sum^{n}_{i=1} a^\dag_{i} a_{i}$ denotes the total particle
number, and $f_{3}^{(1)}(N)=-N+2$ and $f_{3}^{(2)}(N)=N-1$
respectively.

Based on Eq. (\ref{fermu3}), the NSSFR of $U(n)$ group can be
directly obtaioned as
\begin{eqnarray}\label{fermun}
\lambda^{h}_{i}&=&\sum^{n}_{\alpha,\beta=1} [ a^\dag_{\alpha}
{\lambda_{i}}^{\alpha \beta} a_{\beta} f_{n}^{(1)}(N) \nonumber\\&&+
 a^\dag_{\alpha}
{\lambda^{\prime}_{i}}^{\alpha \beta} a_{\beta} f_{n}^{(n-1)}(N)],
\end{eqnarray}
where $\lambda^{\prime}_{i}=U(-\lambda^{*}_{i})U^{\dag}$, the matrix
elements of $U$ are $ U_{m,n+1-m}=(-1)^{m+1}$ and the others are
zeros.
%\begin{eqnarray}
%\lambda^{\prime}_{i}=U(-\lambda^{*}_{i})U^{\dag} \\
%U_{m,n+1-m}=(-1)^{m+1}
%\end{eqnarray}

%From the above discussion, one can notice that the classical
%Schwinger fermion representations of the $SU(n)$ give $n-1$ unitary
%irreducible representations according with the total number from $1$
%to $n-1$. Specially, the representations in the subspace with the
%particles number $N^{f}=m$ and the one with $N^{f}=n-m$ are
%conjugations mutually. And the operator function $f^{(m)}_{n}(N)$
%can select the $m$-th representation. Generally, a high order
%representation of the $U(n)$ Lie algebra can be constructed as
%\begin{eqnarray}\label{sun}
%G^{h}_{i}=\sum^{n-1}_{m=1} \sum^{n,n}_{\alpha=1,\beta=1}
%a^\dag_{\alpha} {G^{(m)}_{i}}^{\alpha \beta} a_{\beta}
%f_{n}^{(m)}(N) \xi_{m},
%\end{eqnarray}

%where for each $m$ either $\xi_{m}=0$ or $1$ and
%$\sum_{m}\xi_{m}\neq 0$ , and $G^{(m)}_{i}$ denotes  $n \times n$
%matrix representations of the $U(n)$ Lie algebra those can be equal
%or not. They are all commutable with the Hamiltonian of
%$n$-dimensional fermionic oscillators in Eq. (\ref{H}).
% When for all $m$, $\xi_{m}=1$ and
%$G^{(m)}_{i}=G_{i}$, Eq. (\ref{sun}) return to the Schwinger
%representation in Eq. (\ref{schrepf}).

\section{Representation of $U(C^{m}_{n})$ group with $n$-fermion \label{ucnm}}

%In the above section, we give a general representation of the $U(n)$
%Lie algebra with $n$ fermions operators. It is shown that the
%construction of the $U(n)$ symmetry of $n$-dimensional fermionic
%oscillators is not unique.

In the $n$-fermion occupation number space, the dimension of the
subspace with $N=m$ is $C^{m}_{n}=n!/m!(n-m)!$. In such a subspace
we can construct the Schwinger fermionic representation of an
$U(C^{m}_{n})$ group. The largest one we can construct is
$U(C^{[\frac{n}{2}]}_{n})$ group, where $[\frac{n}{2}]$ denotes the
integer part of $n/2$.

Let us take the $4$-fermion case as an example. The subspace with
the total number $N=2$ is $\{a^{\dag}_{1}a^{\dag}_{2},
a^{\dag}_{1}a^{\dag}_{3}, a^{\dag}_{1}a^{\dag}_{4},
 a^{\dag}_{2}a^{\dag}_{3}, a^{\dag}_{2}a^{\dag}_{4}, a^{\dag}_{3}a^{\dag}_{4}
\}|vac\rangle$, where $a^{\dag}_{i}$ is the $i$-th fermionic
creation operator. We introduce the notations as $O_{1}=a_{2}a_{1},
O_{2}=a_{3}a_{1}, O_{3}=a_{4}a_{1}, O_{4}=a_{3}a_{2},
O_{5}=a_{4}a_{2}, O_{6}=a_{4}a_{3}$ and
$|i\rangle=O^{\dag}_{i}|vac\rangle$. Then the operators $Q_{ij}$,
which behave as the matrices $e_{ij}$, in the subspace are
\begin{eqnarray}
Q_{ij}=|i\rangle\langle j|=O^{\dag}_{i}|vac\rangle \langle
vac|O_{j}=O^{\dag}_{i}O_{j}f^{(2)}_{4}(N),
\end{eqnarray}
where the selective function $f^{(2)}_{4}(N)=-N^{2}+4N-3$.

The commutation relations among $Q_{ij}$ are the same as those of
$e_{ij}$, i.e.,
$[Q_{ij},Q_{kl}]=\delta_{jk}Q_{il}-\delta_{li}Q_{kj}.$.
%\begin{eqnarray}
%[Q_{ij},Q_{kl}]=\delta_{jk}Q_{il}-\delta_{li}Q_{kj}.
%\end{eqnarray}
All of the 36 $Q_{ij}$'s commute with the total number operator. If
$G_{i}$ denotes the fundamental representation of $i$-th generator
of the $U(6)$ Lie group, then its Schwinger fermionic representation
can be realized as
\begin{eqnarray}
G^{h}_{i}&=&\sum^{6,6}_{\alpha=1,\beta=1} G^{\alpha\beta}_{i}
Q_{\alpha\beta}
\nonumber\\&=&\sum^{6,6}_{\alpha=1,\beta=1}O^{\dag}_{\alpha}
G^{\alpha\beta}_{i} O_{\beta} f^{(2)}_{4}(N).
\end{eqnarray}

%By some calculations, we derive another form of the operators
%\begin{eqnarray}
%Q_{ij}=O^{\dag}_{i}O_{j}g_{(i,j)},
%\end{eqnarray}
%where $g_{i,j}$ is the product of the operators $a_{k}a^{\dag}_{k}$
%over all $k$ with neither $a_{k}$ being included in $O_{j}$ nor
%$a^{\dag}_{k}$ in $O^{\dag}_{i}$.

In general, one can construct the $U(C^{m}_{n})$ Lie algebra using
$n$-fermion operators. To derive the mapping, we first define a set
$n$-dimensional vectors $\zeta=(\zeta_{1},\zeta_{2}...,\zeta_{n})$
where each $\zeta_{i}=0$ or $1$.
%If the first different element of
%two vectors satisfy $\zeta_{j}
%> \zeta^{\prime}_{j}$, we say $\zeta$ is greater than $\zeta^{\prime}$  and denote as  $\zeta >
%\zeta^{\prime}$.
There are $2^{n}$ such vectors, and $C^{m}_{n}$ of them satisfy
$\sum^{n}_{i=1} \zeta_{i}=m$ for a fixed integer m, where $0 \leq m
\leq n$. Label the $C^{m}_{n}$ vectors in the descending order, e.g.
the first one is
\begin{eqnarray}
\zeta^{(1)}=(\zeta^{(1)}_{1}=1,...,\zeta^{(1)}_{m}=1,
\zeta^{(1)}_{m+1}=0,..., \zeta^{(1)}_{n}=0).
\end{eqnarray}
Due to $\zeta^{(i)}$, we can define a set of operators as
\begin{eqnarray}
O_{i}=a_{n}^{\zeta^{(i)}_{n}}a_{n-1}^{\zeta^{(i)}_{n-1}}...a_{1}^{\zeta^{(i)}_{1}}.
\end{eqnarray}
The element operators are similarly defined as
$Q_{ij}=O^{\dag}_{i}O_{j}f^{(m)}_{n}(N)$.
%\begin{eqnarray}
%Q_{ij}=O^{\dag}_{i}O_{j}g_{(i,j)},
%\end{eqnarray}
%where $g_{(i,j)}$ is the product of the operators
%$a_{k}a^{\dag}_{k}$ over all $k$ with both the degrees of $a_{k}$
%and $a^{\dag}_{k}$ in $O^{\dag}_{i}O_{j}$ are zero.
Their commutation relations satisfy
$[Q_{ij},Q_{kl}]=\delta_{jk}Q_{il}-\delta_{li}Q_{kj}$.
%\begin{eqnarray}
%[Q_{ij},Q_{kl}]=\delta_{jk}Q_{il}-\delta_{li}Q_{kj}.
%\end{eqnarray}
Then the Schwinger fermionic representations of the $U(C^{m}_{n})$
group can be realized by
\begin{eqnarray} \label{su6of4}
G^{m}_{i}=\sum_{\alpha,\beta} G_{i}^{\alpha \beta}Q^{m}_{\alpha
\beta},
\end{eqnarray}
where $G_{i}$ is the $i$-th $C^{m}_{n} \times C^{m}_{n}$ matrix
generator of the $U(C^{m}_{n})$ group, and $G_{i}^{\alpha \beta}$ is
its element in $(\alpha,\beta)$ position.

Let $\bar{m}=n-m$, $C^{\bar{m}}_{n}=C^{m}_{n}$. The subspaces with
$N=m$ and $N=\bar{m}$ are mutually conjugate. The linear
combinations of the operators $Q^{\bar{m}}_{ij}$
\begin{eqnarray}
G^{\bar{m}}_{i}=\sum_{\alpha,\beta} G_{i}^{\alpha
\beta}Q^{\bar{m}}_{\alpha \beta}
\end{eqnarray}
also obey the commutation relations of $U(C^{m}_{n})$ algebra.
Therefore, when $\bar{m} \neq m$, the more general representation of
$U(C^{m}_{n})$ are given by
\begin{eqnarray} \label{universal}
G^{h}_{i}=\sum_{\alpha \beta} G^{\alpha \beta}_{i} Q^{(m)}_{\alpha
\beta} \xi_{-} + \sum_{\alpha \beta} G^{\prime \alpha \beta}_{i}
Q^{(\bar{m})}_{\alpha \beta} \xi_{+},
\end{eqnarray}
where $\xi _{\pm}=0$ or $1$, $\xi_{+}+\xi_{-} \neq 0$, $G$ and
$G^{\prime}$ denotes two $C^{m}_{n} \times C^{m}_{n}$ matrix
representations of the $U(C^{m}_{n})$ group.

%The maximum of $C^{m}_{n}$, $Max(C^{m}_{n})=C^{[\frac{n}{2}]}_{n}$,
%where $[\frac{n}{2}]$ denotes the integer part of $\frac{n}{2}$.  By
%using $n$ sets of fermionic operators, in the above way, the largest
%special unitary group is the $U(C^{[\frac{n}{2}]}_{n})$. When $n$ is
%even, $[\frac{n}{2}]=\frac{n}{2}$, the construction is
%\begin{eqnarray}\label{neven}
%G^{h}_{i}=\sum_{\alpha \beta} G^{\alpha \beta}_{i}
%Q^{(\frac{n}{2})}_{\alpha \beta}.
%\end{eqnarray}
%When $n$ is odd, $[\frac{n}{2}]=\frac{n-1}{2}$ and
%$\bar{\frac{n-1}{2}}=\frac{n+1}{2}$, the largest construction is
%\begin{eqnarray} \label{nodd}
%G^{h}_{i}=\sum_{\alpha \beta} G^{\alpha \beta}_{i}
%Q^{(\frac{n-1}{2})}_{\alpha \beta} \xi_{-} + \sum_{\alpha \beta}
%G^{\prime \alpha \beta}_{i} Q^{(\frac{n+1}{2})}_{\alpha \beta}
%\xi_{+}.
%\end{eqnarray}

%When $n=2$, $C^{1}_{2}=2$, one can choose $G_{i} =
%\frac{\sigma_{i}}{2}$ ($\sigma_{i}$ is Pauli matrices, $i=1,2,3$) ,
%and Eq. (\ref{neven}) gives the usual Schwinger fermion
%representations of the $SU(2)$ Lie algebra.

%When $n=3$, $C^{1}_{3}=C^{2}_{3}=3$. Let
%$G_{i}=G^{\prime}_{i}=\lambda_{i}$ and $\xi_{\pm}=1$, Eq.
%(\ref{nodd}) returns to the higher order representation in Eq.
%(\ref{hordersu3}). If choose $G_{i}=\lambda_{i}$ and
%$G^{\prime}_{i}=\lambda^{\prime}_{i}$, Eq. (\ref{nodd}) gives the
%Schwinger representations.

%When $n=4$, the result of Eq. (\ref{neven}) is the same as the one
%in Eq. (\ref{su6of4}).

\section{conclusion \label{conclu}}

In conclusion, we have studied the non-standard Schwinger fermionic
representation of the unitary group by using $n$-fermion operators.
we found that the Schwinger fermionic representation of the $U(n)$
group is not unique when $n\ge 3$. In general, based on $n$-fermion
operators, the NSSFR of the $U(n)$ group can be established in a
uniform approach, where all the generators commute with the total
number operators. The Schwinger fermionic representation of
$U(C^{m}_{n})$ group is also discussed.

We thank S. W. Hu for his valuable discussions. This work is
supported in part by NSF of China (Grant No. 10575053 and No.
10605013) and Program for New Century Excellent Talents in
University.

%In the investigation of the connection between $3$-fermion and the
%$SU(3)$ Lie algebra, we find a universal construction of the $U(n)$
%Lie algebra using n fermions. The result indicates that the $U(n)$
%symmetry of the $n$-dimensions fermionic oscillators has multiform
%descriptions. Sequentially, $4$ fermions can be used to construct a
%representation of the $SU(6)$ Lie algebra. Generally, we give a
%construction of the $U(C^{m}_{n})$ Lie algebra with $n$ fermions,
%the largest one of which is the $U(C^{[\frac{n}{2}]}_{n})$ case.
%These constructions all conserve the total number operators. Given
%proper parameters, they can return to the usual Schwinger fermion
%representations.

%\begin{acknowledgments}
%We thank S. W. Hu for his valuable discussions. This work is
%supported in part by NSF of China (Grant No. 10575053 and No.
%10605013) and Program for New Century Excellent Talents in
%University.
%\end{acknowledgments}

\end{document}